\documentclass[prc,floatfix,nofootinbib,showpacs,showkeys,reprint]{revtex4-1}
\usepackage[utf8]{inputenc}
\usepackage[T1]{fontenc}
\usepackage{amsmath}
\usepackage{amsfonts}
\usepackage{amssymb}
\usepackage{graphicx}
\usepackage{xcolor}
\usepackage{gensymb}

\newcommand{\JLU}{Institut f\"ur Theoretische Physik, Universit\"at Giessen, Giessen, Germany}
\newcommand{\HFHF}{Helmholtz Research Academy Hesse for FAIR (HFHF), Campus Giessen, Giessen, Germany}

\begin{document}

\title{Dilepton Decay of Low-mass $\rho$ Mesons }

\author{K. Gallmeister}
\email[Contact e-mail: ]{kai.gallmeister@theo.physik.uni-giessen.de}
\affiliation{\JLU}
\affiliation{\HFHF}

\author{U. Mosel}
\affiliation{\JLU}
\affiliation{\HFHF}

\author{L. von Smekal}
\affiliation{\JLU}
\affiliation{\HFHF}

\begin{abstract}
	The HADES collaboration has extracted dilepton mass spectra for $\pi^-$ induced reactions on the proton from a comparison of data taken on C and CH$_2$ targets. The spectra were interpreted in terms of different versions of vector meson dominance. Here we present results obtained from the theory and generator GiBUU. We first check
	the subtraction procedure used and then discuss the obtained mass spectra for the proton target. We point out that any conclusions on the version of VMD requires the knowledge of the $\rho$ spectral function in the interesting mass region.
\end{abstract}

\maketitle

\section{Introduction}
The radiative decay of nucleon resonances is governed by electromagnetic transition form factors which take the finite extension of the resonance and the nucleon into account. A very successful description of such  photon-hadron couplings and  has been achieved within the vector-meson-dominance (VMD) model \cite{Nambu:1962zz}. In this model the coupling takes place through an intermediate $\rho$ meson. As discussed in some detail in \cite{OConnell:1995nse} there are two versions of VMD. In the more widely used version VMD2 (in the notation of \cite{OConnell:1995nse}) the coupling of the photon takes place only through the $\rho$ meson, whereas in version VMD1 the coupling amplitude employs in addition a coherently added term in which the photon couples directly to the hadron. Both methods are fully equivalent if certain relations between the coupling constants involved are met \cite{OConnell:1995nse}. Since these relations in nature hold only approximately the HADES collaboration has recently tried to find experimental signatures for one or the other VMD version \cite{HADES:2022vus}.

The two VMD versions differ in their predicted dilepton invariant mass ($M_{e^+e^-}$) distribution mainly at small masses below the $2 m_\pi$ threshold where the pion electromagnetic formfactor is not accessible. One possibility to explore this low mass range is thus given by the Dalitz decay of a nucleon resonance $N^* \to N e^+e^-$; this decay is governed by the electromagnetic transition formfactor in the time-like region. A conclusive comparison of data with the VMD versions should be possible if the $\rho$-spectral function in the mass range used for the comparison is known.

Therefore, merging the experimental information on $\rho$ production in the $\pi^- + p \to \rho + n$ reaction from Ref.~\cite{HADES:2020kce} with measured dilepton yields in the reaction $\pi^- + p \to e^+e^- + n$ from Ref.~\cite{HADES:2022vus} seems to offer an interesting possibility to access also the low-mass region and to explore the validity of VMD there.

The recent HADES experiment aims to determine the dilepton yield in the reaction $\pi^- p$ at dilepton invariant masses of about 100\,MeV to 400\,MeV, i.e.~around and below the $2\pi$ threshold \cite{HADES:2022vus}. The experiment was performed at an incoming pion momentum of $p_\pi = 0.69$\,GeV, corresponding to $\sqrt s = 1.49$\,GeV. It did not directly use a proton target, but instead obtained data for CH$_2$ and (with lower statistics) for C. A comparison of both then leads to the dilepton mass spectrum for H. Invariant mass cuts, assuming a quasi-free reaction process, are used to isolate the $n + e^+e^-$ final state.

The further analysis uses results of an earlier publication  \cite{HADES:2020kce} on a measurement and analysis of 2$\pi$ production in the $\pi^- + p$ reaction.
 A partial wave analysis (PWA) of these results employing the Bonn-Gatchina $K$-matrix model \cite{BnGa,Anisovich:2004zz} then led to the cross section for $\rho$ production primarily through the D$_{13}$ N(1520) resonance.

The data obtained in \cite{HADES:2022vus} do not show the strong rise ($\propto 1/M_{e^+e^-}^3$) towards small dilepton invariant masses contained in the VMD2 model \cite{Larionov:2020fnu}. On the other hand, a good fit to the data could be obtained by using VMD1 and  fitting the relative strength of the two components with a free parameter.

For such a comparison of the VMD versions the $\rho$ spectral function has to be known in the region of interest.
Crucial for this comparison is, therefore, the continuation of the $\rho$ spectral function from an energy range, where it has been measured by the $2 \pi$ decay, to lower masses. The latter is influenced by the decay N$(1520) \to \rho + n$, i.e.~by a hadronic interaction vertex independent of the electromagnetic coupling of the $\rho$ to the virtual photon. In VMD1 an additional assumption about the electromagnetic transition formfactor is required.

The purpose of this present paper is twofold. First, we investigate the steps leading from the actual measurements, which were performed on heavier targets (C, CH$_2$), to the spectrum on a proton. Second, we have a closer look at the conditions necessary to decide between VMD1 and VMD2.

\section{Model}
We describe the reactions $\pi^- + C$ and $\pi^- + p$ within the quantum-kinetic GiBUU theory and event generator. Both the theoretical foundations as well as all the elementary reaction input and the numerical algorithms are described in detail in Ref.~\cite{Buss:2011mx}. The source code used for the present calculations can be obtained from \cite{gibuu}. All the special equations for dilepton production are given in Ref.~\cite{Larionov:2020fnu}. As described there
we use the VMD2 variant for the dilepton-decay of a $\rho$ meson.

Essential input for the calculations are the properties of the hadronic interaction channels; for the problem at hand the relevant channel is $\pi^- + p \to n + \rho$. In the present version of GiBUU these are taken from the partial wave analysis of Manley and Saleski \cite{Manley:1992yb}.\footnote{The HADES analysis of the 2$\pi$ cross sections in \cite{HADES:2020kce} used the Bonn-Gatchina analysis which leads to somewhat different decay widths.}
Contrary to the non-relativistic spectral function employed there in the present calculations we use a relativistic form which -- when integrated over $m$ -- is normalized to 1 for constant widths
\begin{equation}
	A(m) = \frac{2}{\pi}\,\frac{m^2 \Gamma(m)}{(m^2 - M^2)^2 + m^2\Gamma(m)^2 }~.
\end{equation}
Here $M$ stands for the peak mass of either the $\rho$ meson or the N(1520) resonance. The width $\Gamma$ is either that equal to the sum of dilepton and pion decay widths for the $\rho$ meson or that of the nucleon resonance. We use this Breit-Wigner form for the spectral function of the $\rho$ meson also below the $2\pi$ threshold even though the physical character of this meson becomes doubtful in this mass range.

In the Manley-Saleski analysis the decay width of a nucleon resonance $N^*$ with mass $M$ into $N \rho$ is given by the product of the phase-space element with a Blatt-Weisskopf (BW) formfactor $B_l$,
\begin{equation}    \label{Gamma*}
	\Gamma^*(W,m) = \Gamma^*_0 \, \frac{q_\rho(W,m)}{W} B_l^2(q_\rho(W,m)\, R)~.
\end{equation}
Here $q_\rho$ is the c.m.~momentum of the decay products,
\begin{align}
	q_\rho(W,m)
	&= \\
	\frac{1}{2W}& \sqrt{ (W^2 - (M_N + m)^2)(W^2-(M_N-m)^2)},\nonumber
\end{align}
$W$ is the invariant mass of the decaying $N^*$, $m$ is the running mass of the $\rho$ meson and $M_N$ is that of the nucleon. The BW formfactor regularizes the width at large $m$. It does not, however, limit the high-momentum (low-mass) behavior.

\begin{figure}
	\centering
	\includegraphics[width=0.7\linewidth]{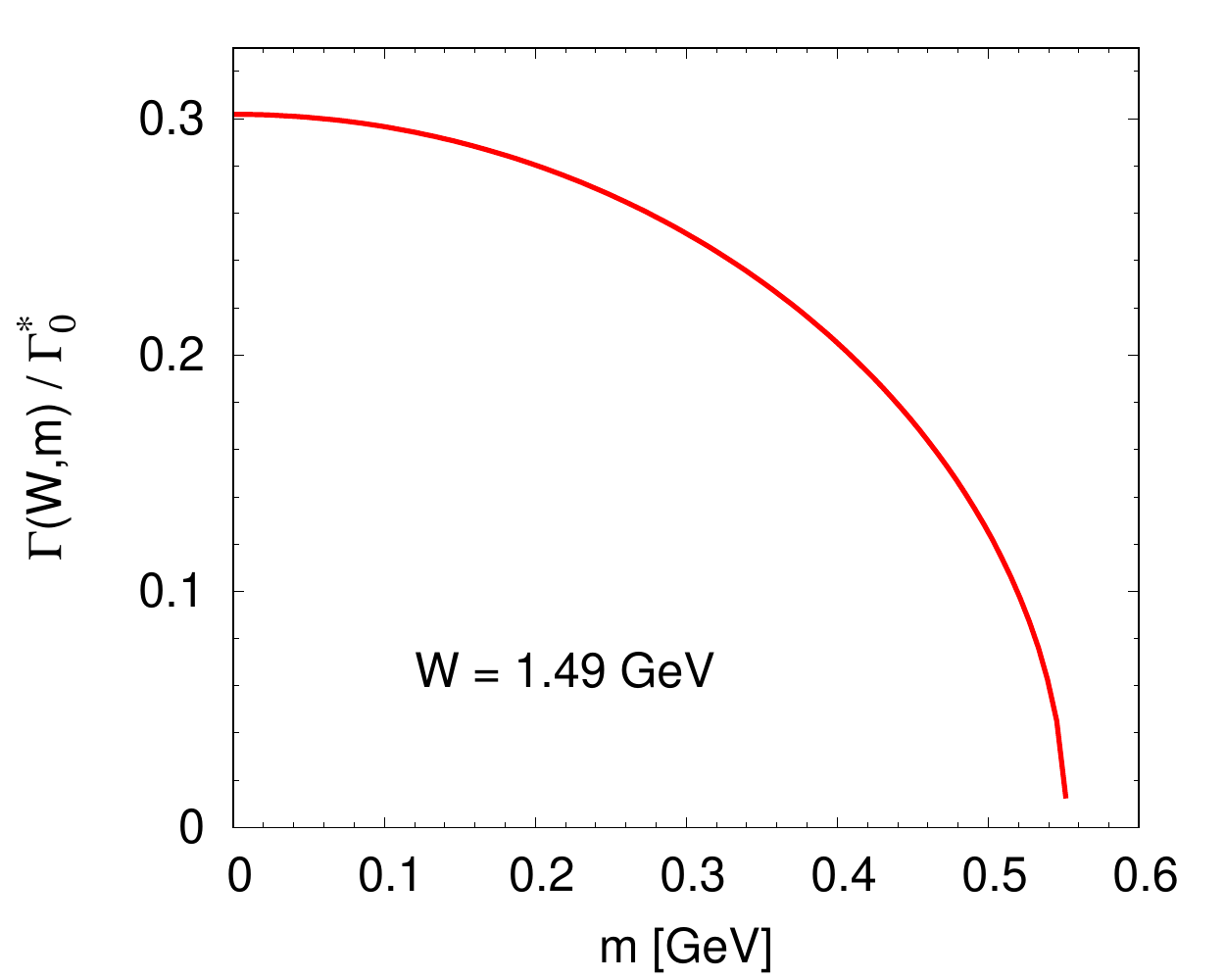}
	\caption{Decay width of Eq.~(\ref{Gamma*}) for the decay $N(1520) \to N \rho$ as a function of the $\rho$ mass $m$ at fixed $W=1.49$ GeV.}
	\label{fig:rhomomentum}
\end{figure}

 In Fig.~\ref{fig:rhomomentum} we show the width for the decay $N(1520) \to \rho N$ at the invariant mass at which the HADES experiment was performed.
Being dominated by phase-space this width obviously favors the production of high-momentum, very-low-mass $\rho$ mesons. This is in contrast to usually used formfactors which are used to mimick the finite extension of hadrons and thus cut off the high-momentum parts of a transition.

The Manley-Saleski analysis does not describe explicitly the secondary decay of primary decay daughters. For the present calculations we use the expression \cite{Post:2003hu}
\begin{eqnarray}   \label{rho-width}
	\Gamma_{\rho \to \pi^+\pi^-}(m) &=& \Gamma_0 \, \frac{M_\rho^2}{m^2} \frac{q_\pi^3(m)}{q_\pi^3(M_\rho)} \frac{1 + [q_\pi(M_\rho)R]^2}{1 + [q_\pi(m)R]^2}  \nonumber \\
	& & \times \Theta(m - 2m_\pi)  ~,
\end{eqnarray}
with
\begin{equation}
	q_\pi(m) = \frac{1}{2} \sqrt{m^2 - 4m_\pi^2} ~,
\end{equation}
being the c.m.~momentum of the pions.\footnote{Note that this definition of the $\rho$ decay width from \cite{Post:2003hu} differs by a factor $M_\rho/m$ from the one used in \cite{Larionov:2020fnu}. For the dilepton spectra from heavy-ion collisions this difference is irrelevant.}
Here $m$ is the 'running mass' of the $\rho$ meson and $M_\rho$ is its nominal mass at the peak of the spectral function.  Finally, the dilepton decay width is given by
\begin{eqnarray}    \label{VMD2}
	\Gamma_{\rho \to e^+e^- }(m) &=& C_\rho \frac{M_\rho^4}{m^3} (1 + 2m_e^2/m^2)\sqrt{1 - 4m_e^2/m^2}  \nonumber \\
	& & \mbox{} \times \Theta(m - 2m_e)~.
\end{eqnarray}
Here $m_e$ is the electron mass and $C = 9.078 \times 10^{-6}$ is an empirical coupling constant \cite{Larionov:2020fnu}.

\section{Results}
\begin{figure}	\centering
	\includegraphics[width=0.8\linewidth]{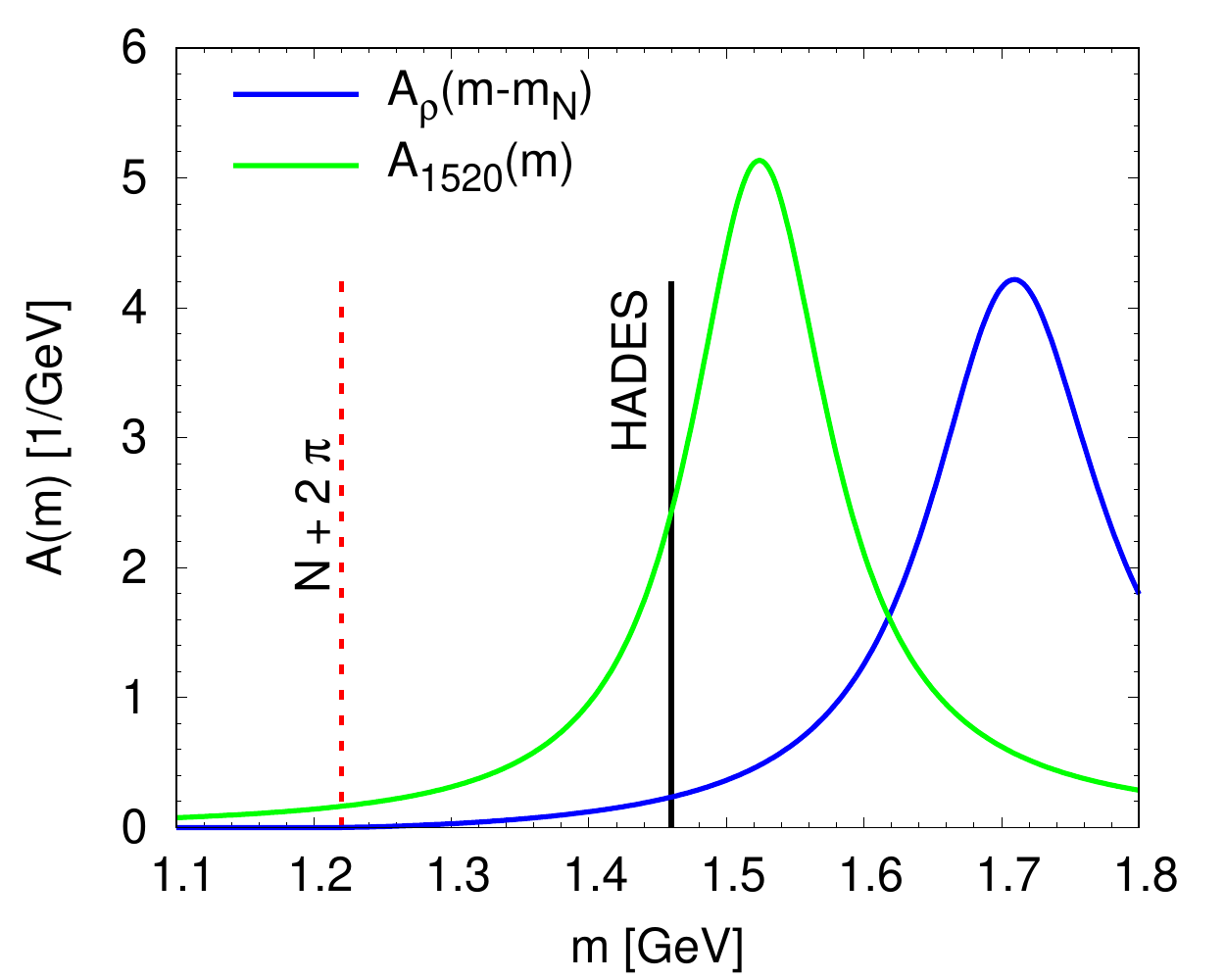}
	\caption{The spectral functions of the N(1520) resonance and the $\rho$ meson, the latter shifted by the nucleon mass. The vertical line at $m = 1.46$\,GeV shows the energy of the HADES experiment; the line at $m = 1.22$\,GeV denotes the $2\pi$ threshold of the $\rho$ meson.}
	\label{fig:n1520rhon}
\end{figure}

We start our discussion with pointing out that the beam energy in this experiment is so low that only the low-mass tail of the $\rho$ meson is populated in the reaction $\pi^- + p \to n + \rho$ and even for the N(1520) the full peak cannot be reached. This is illustrated in Fig.~\ref{fig:n1520rhon} showing the spectral functions for the N(1520) resonance and the $\rho$ meson, the latter shifted by the nucleon mass.

The experiment was run at an $\pi N$ invariant mass of 1.46\,GeV (indicated by a vertical bar in Fig.~\ref{fig:n1520rhon}). For the present experiment even the very low mass region below the $2\pi$ threshold of the $\rho$ meson ($m < 1.22$\,GeV) is essential because it is here that the predictions of VMD1 and VMD2 differ most in the analysis of \cite{HADES:2022vus}. Fig.~\ref{fig:n1520rhon} shows that this region lies in the extremely low low-mass tail of the $\rho$.
Here one has to keep in mind that  the Breit-Wigner distribution for the $\rho$ meson emerges from a Taylor expansion of the $\pi \pi$ scattering amplitudes around the peak mass and thus becomes the less reliable the farther the masses are removed from the maximum.

\subsection{$\pi^- p$ cross sections}

\begin{figure}
	\centering
	\includegraphics[width=0.9\linewidth]{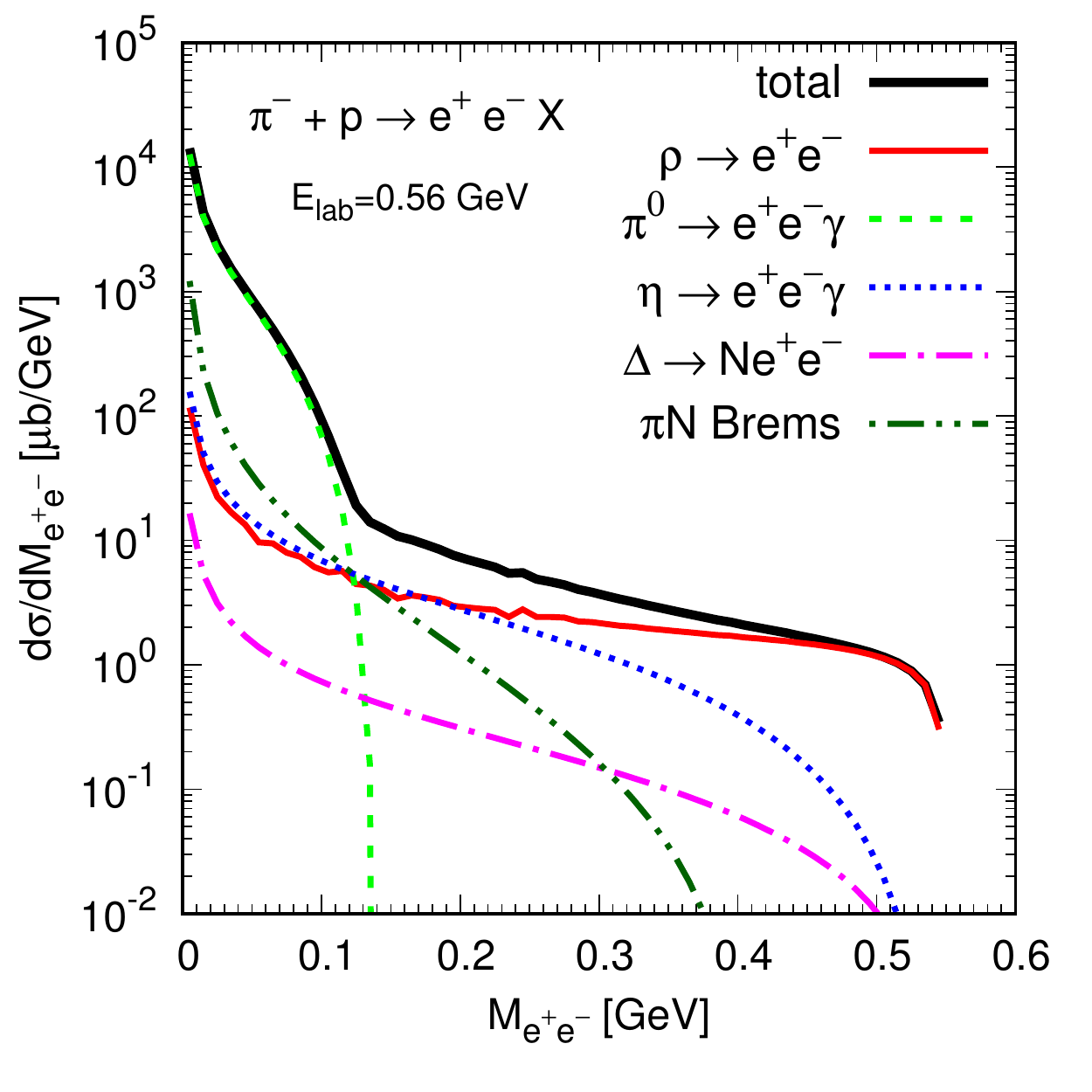}
	\caption{Invariant mass spectrum of dileptons in the reaction $\pi^- + p$ at 0.56\,GeV lab energy. The individual contributions are indicated in the figure.}
	\label{fig:pipspectrum}
\end{figure}

In Fig.~\ref{fig:pipspectrum} we show for reference the calculated various dilepton contributions for the reaction $\pi^- + p \rightarrow n + e^+e^-$, with the proton isolated and at rest.
In the relevant mass range, starting at the upper end of the $\pi^0$ Dalitz decay at around 0.14\,GeV and extending to about 0.4\,GeV, well above the two-pion threshold, the dominant contributions are the radiative $\rho$ decay and the $\eta$ Dalitz decay; at the lowest masses in this range also $\pi N$ bremsstrahlung is non-negligible.

\begin{figure}
	\centering
	\includegraphics[width=0.9\linewidth]{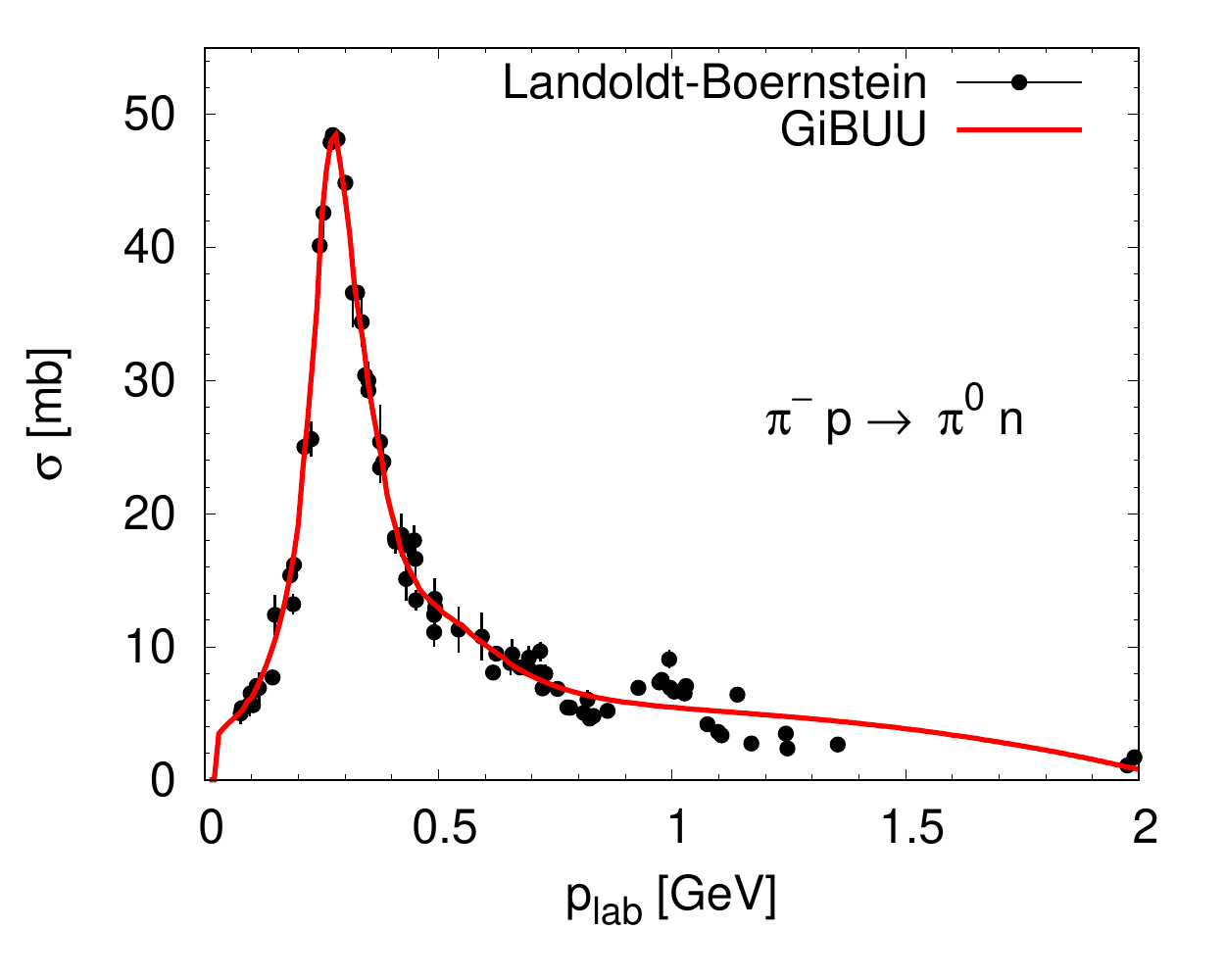}
	\caption{Inclusive cross section for $\pi^- + p \to \pi^0 + n$ as a function of the incoming pion's lab momentum. Data can be found in \cite{Schopper:1988vrx}.}
	\label{fig:pi-p-n-pi0}
\end{figure}

We have checked the absolute magnitude of these cross sections by comparing the cross section for $\pi^- + p \to \pi^0 + n$ with experiment \cite{Schopper:1988vrx} (see Fig.~\ref{fig:pi-p-n-pi0}); excellent agreement is reached.
This fixes the height of the $\pi^0$ Dalitz peak to the correct value.

\subsection{Extraction of $\pi^- p$ cross sections\\ in the HADES experiment}

The HADES experiment was not run on a proton target, but instead used an approximate subtraction procedure based on measured missing mass distributions for CH$_2$ and C to isolate the proton contribution \cite{HADES:2020kce}. By cuts  on these distributions, furthermore, the experiment has tried to identify the particular contribution $\pi^- + p \to n + \rho \to n + e^+e^-$,  assuming a quasifree interaction with the protons in the target nucleus.

\begin{figure}
	\centering
	\includegraphics[width=0.9\linewidth]{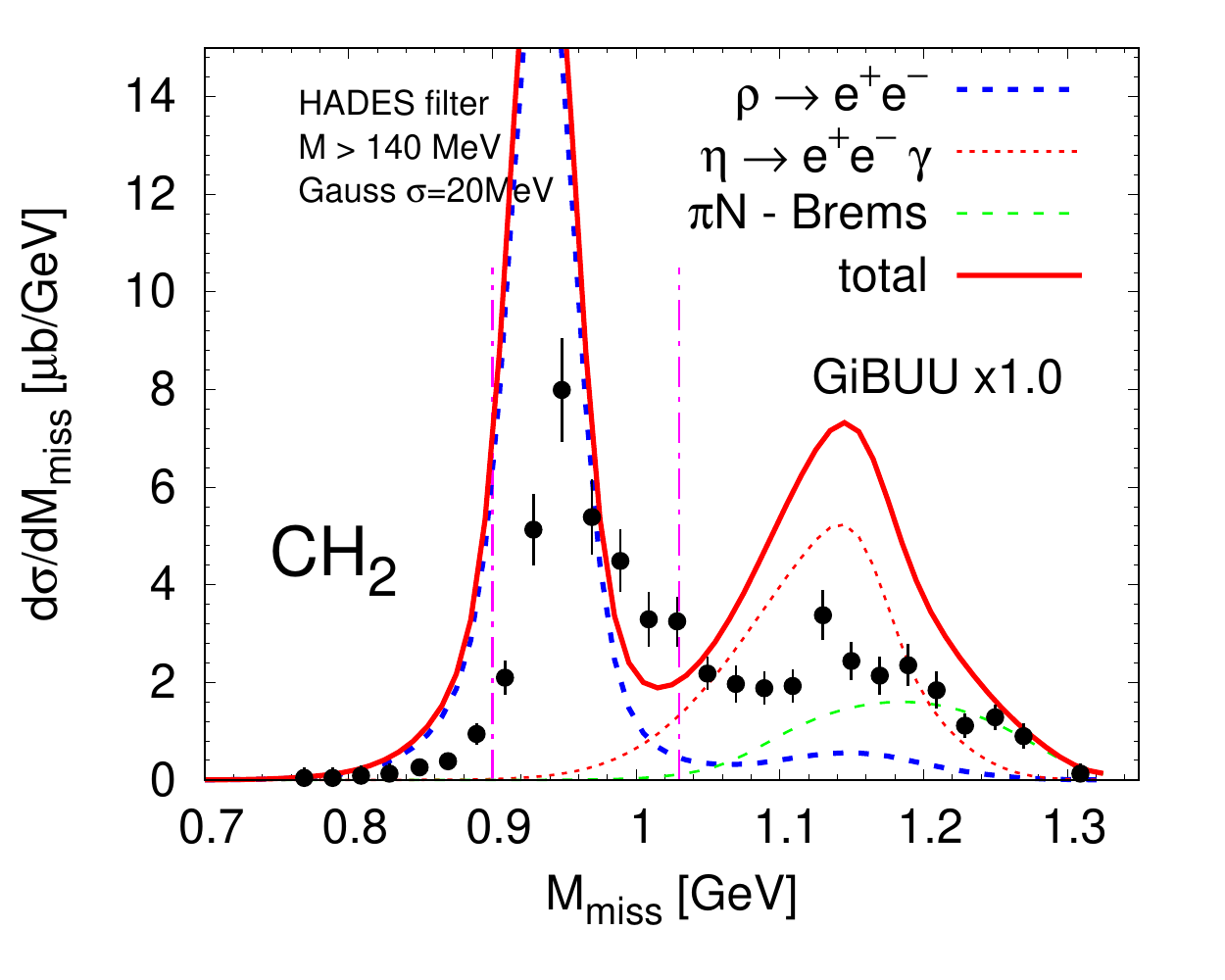}
	\caption{Missing mass distribution for $\pi^- + CH_2$ for events with an invariant mass larger than 140\,MeV. Shown are the contribution of the $\rho$ meson (dashed, blue curve), of the $\eta$ meson (dotted, red), the $\pi N$ bremsstrahlung and the sum of all contributions (solid, red). The results of the calculation include the effects of angular cuts according to the HADES acceptance. The data points are from \cite{HADES:2022vus}, they are efficiency, but not acceptance corrected. The two vertical lines show the positions of the missing-mass cuts employed by the experiment.}
	\label{fig:missmassmdatach2smear10}
\end{figure}

To check this procedure we first show in Fig.~\ref{fig:missmassmdatach2smear10} a comparison of the missing mass distribution obtained with GiBUU for the CH$_2$ target. The results were obtained by using the experimental HADES acceptance cuts\footnote{The HADES acceptance cuts used were: $\theta_i = 18\degree - 85\degree,p_i > 100$\,MeV, opening angle between leptons $> 9 \degree$.}.
In agreement with the data the missing mass distribution shows two clearly distinct peaks, one centered at the free neutron mass and another one at 1.15\,GeV, representing the $\eta$ Dalitz-decay contribution. Both peaks also show up in the experimental distribution. There is, however, a significant disagreement in absolute height of the cross section; the GiBUU cross section is larger than then experimental values by about a factor 3. The missing mass distribution shows a clear peak at the neutron mass $M_n = 0.938$\,GeV while the result of a simulation shown in Fig.~ 1 of Ref.~\cite{HADES:2022vus} peaks at a higher mass. Part of this discrepancy in the mass is due to a detector-efficiency related smearing. The overall magnitude of the cross sections is related to additional acceptance limitations of the HADES detector. Unfortunately, a filter routine that takes these acceptance properties into account is still not available.

\begin{figure}
	\centering
	\includegraphics[width=0.9\linewidth]{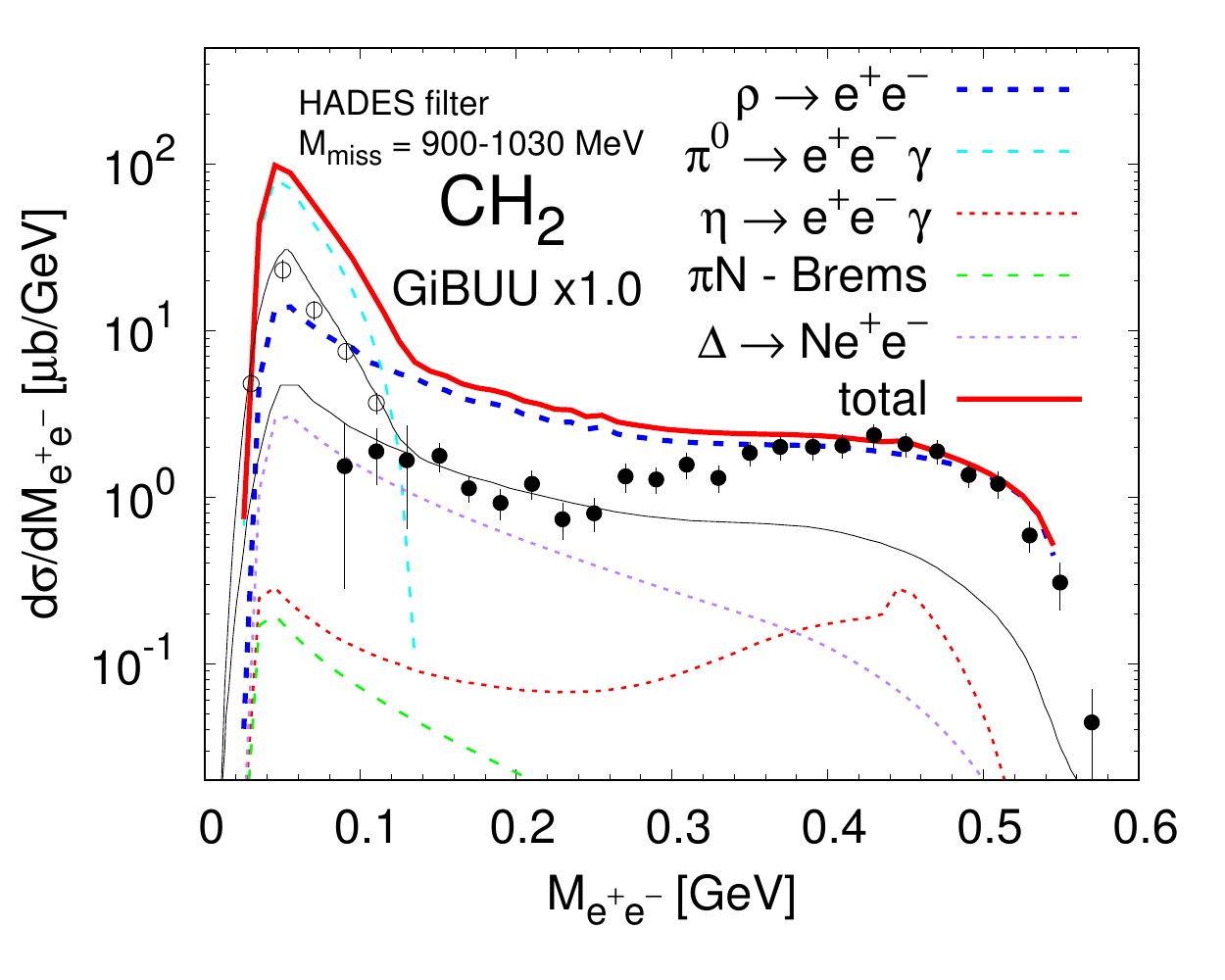}
	\caption{Dilepton invariant mass distribution for $\pi^- + CH_2$. Shown are the contribution of the $\rho$ meson (dashed, blue curve), of the $\eta$ meson (dotted, red), the $\pi N$ bremsstrahlung and the sum of all contributions (solid, red). The results of the calculation include the effects of angular cuts according to the HADES acceptance. The data points are from \cite{HADES:2022vus}, they are corrected for detector efficiencies, but restricted to the HADES acceptance cuts given above. Thin black lines show results from \cite{HADES:2022vus}.}
	\label{fig:invmasCH2}
\end{figure}

In the GiBUU calculations the CH$_2$ cross section is simply given by the sum of cross sections for C and for $2 \times H$. In Fig.~\ref{fig:invmasCH2} we show first the dilepton spectrum for CH$_2$ since this spectrum is directly measured and does not suffer from possible inaccuracies in the extraction of the proton cross section. The calculation describes the data very well for masses above about 0.4 GeV, i.e.~above the $2\pi$ threshold. Going to smaller masses it becomes larger than the data by about a factor 3. Surprising is the fact that this overestimate also shows up in the $\pi^0$ Dalitz peak region since the elementary charge exchange cross section for $\pi^- + p \to n + \pi^0$ is described very well.

The HADES analysis has then used a similar, but statistically less constrained, distribution for a C target to deduce a factor of about 2.9 between the C and p targets \cite{HADES:2022vus}.  This factor is used to normalize the final data on the proton target.

\begin{figure}
\centering
\includegraphics[width=0.9\linewidth]{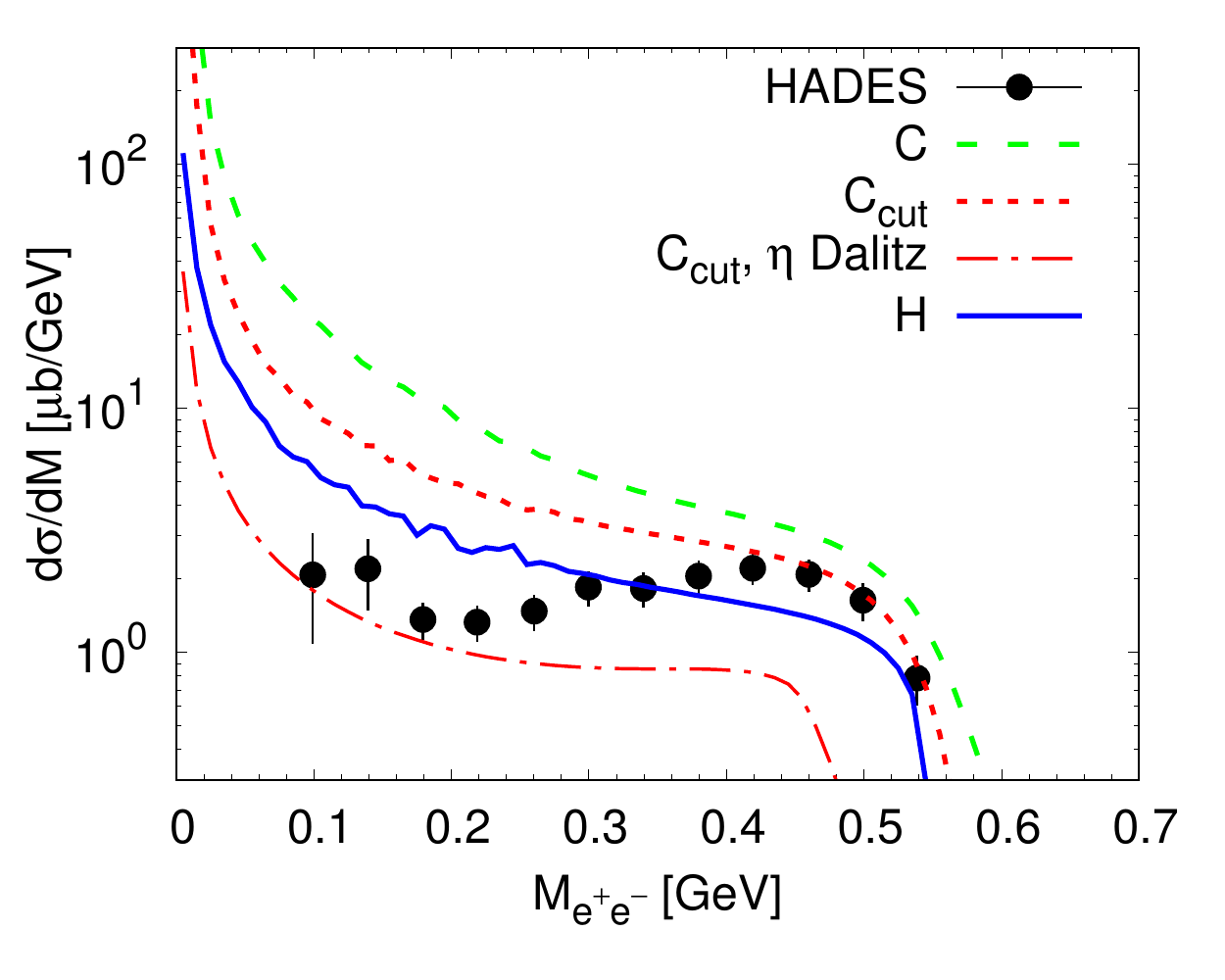}
\caption{Invariant mass spectrum of dileptons for the $\pi^- C$ reaction at 0.56\,GeV (dashed, green curve); the $\pi^0$ Dalitz decay contribution is subtracted. Shown also is the mass distribution after a missing mass cut for the final state $n + e^+e^-$ (dashed red curve, labeled C$_{\rm cut}$) as well as the same for a proton target (blue, solid curve, labeled H). The black data points are taken from \cite{HADES:2020kce}; they are acceptance corrected. The lowest dashed-dotted curve gives the contribution from $\eta$ Dalitz decay after the invariant mass cut.}
\label{fig:invmassdileptons}
\end{figure}

In Fig.~\ref{fig:invmassdileptons} we show the dilepton invariant mass distribution for a $^{12}$C target together with that for H. The cross section is described quite well for the higher masses but overshoots the measured values for masses below about 0.4\,GeV. The final state interactions have only a very small influence justifying the assumption of a quasifree reaction process. The spectra are obtained after performing the invariant mass cuts as in the experiment. It is seen that the resulting distribution, given by the dashed  red line, is higher than the distribution obtained from the H target (blue solid curve) by about a factor of 1.75, significantly smaller than the factor 2.9 used in the HADES analysis. That this factor is less than the number of protons in the target (6) is due to initial state interactions. That it is also lower than the factor obtained in the HADES analysis we attribute to the fact that the results for C shown here contain both the Fermi-motion of the nucleons and their binding energy. It can also be seen that some $\eta$ Dalitz decay contribution survives. While it is well suppressed at the higher masses, below about 200\,MeV it becomes comparable in magnitude to the data. We speculate that the slight upwards trend seen in the data at the lowest masses could be due to this $\eta$ contamination.

Summarizing the results discussed so far we find that the method used by the HADES collaboration to isolate the $\rho$ contribution to the dilepton spectrum works reasonably well with some $\eta$ contribution remaining at the smallest masses. Any detailed comparison, however, requires the knowledge of the HADES acceptance and efficiency.

\subsection{Dilepton invariant mass distribution for H}
After having discussed the preparation of the dilepton spectrum for H, starting from data for $^{12}$C and CH$_2$, we now discuss the mass-dependence of the dilepton spectrum. As shown by the blue, solid curve in Fig.~\ref{fig:invmassdileptons}  the spectrum for H is on average quite well described. However, it is somewhat smaller (by about 30\%) for the masses between 0.3 and 0.5\,GeV and larger (by about 100\%) for the masses around 0.2\,GeV.

\begin{figure}
	\centering
	\includegraphics[width=0.9\linewidth]{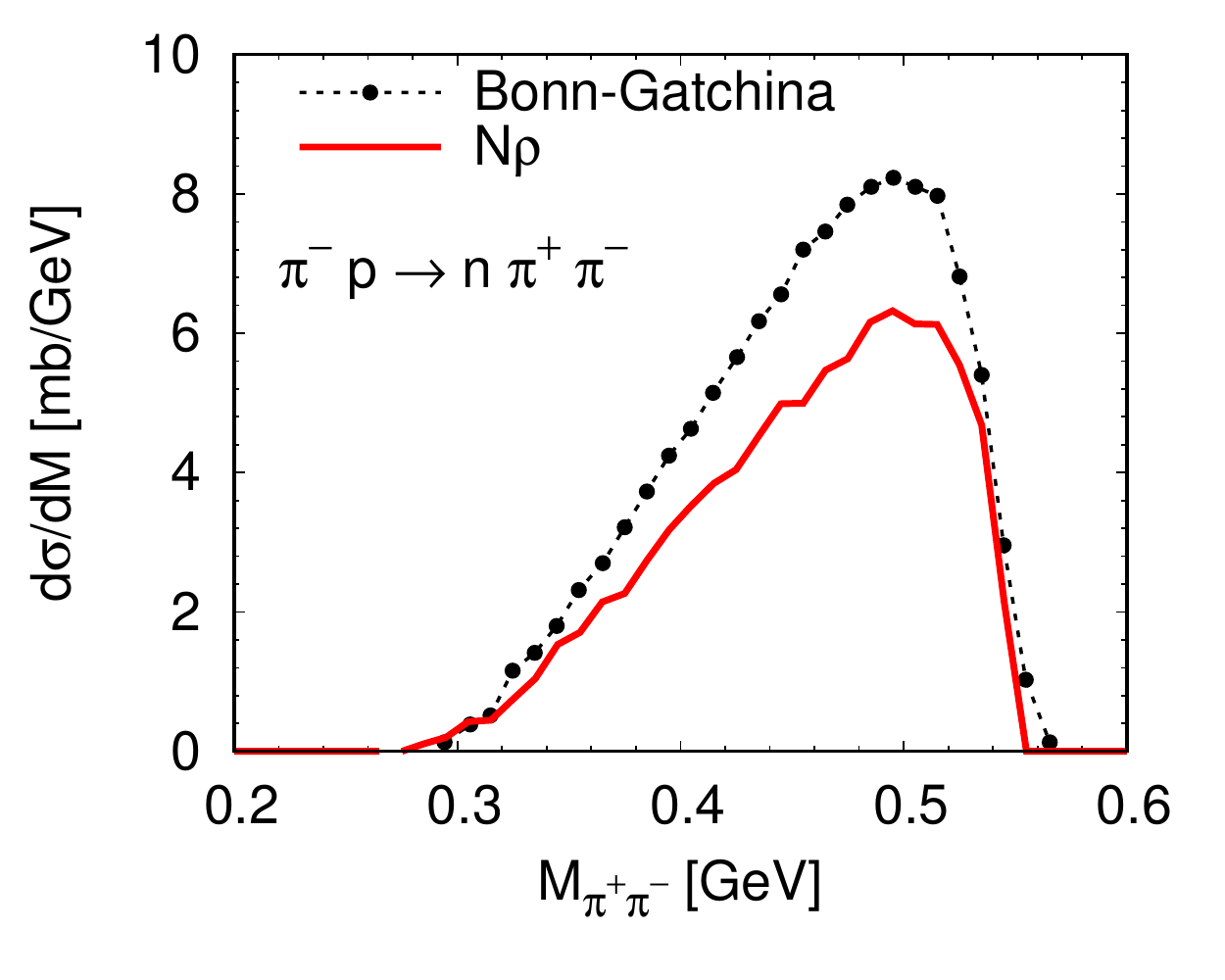}
	\caption{The $\rho$ meson mass distribution for the $\pi^- + p$ reaction. Black points give the $\rho$ contribution extracted from the $2\pi$ data measured by HADES \cite{HADES:2020kce} by means of the Bonn-Gatchina partial wave analysis. The solid (red) curve gives the contribution obtained from a GiBUU calculation.}
	\label{fig:invmassonlyrho}
\end{figure}

 In a model that links an observed dilepton yield to the $\rho$ meson (as in VMD2, but note that also in VMD1 the $\rho$ strength enters) the overall magnitude of the dilepton yield is directly proportional to the number of $\rho$ mesons produced. Fig.~\ref{fig:invmassonlyrho} shows that the extracted experimental mass distribution lies by about a factor of 1.3 above the GiBUU calculation.
This experimental result depends on the particular PWA used to extract the $\rho$-contribution from the measured $2\pi$ cross sections (Bonn-Gatchina for the HADES data, Manley for the GiBUU calculation). Tuning the $\rho$ production cross section to the Bonn-Gatchina values used by HADES the dilepton invariant mass spectrum would improve the description of the data above the $2\pi$ threshold while at the same time increasing the discrepancy at smaller masses.

Note that in the region where the $\rho$ spectral function is known, i.e.~above the $2\pi$ threshold, the agreement of the calculation with the data is quite reasonable. A comparison with the experimental results shown in Fig.~2b of \cite{HADES:2022vus} shows that the tuned VMD1 calculation there essentially agrees with the results obtained here for H.

\section{Discussion}
The very low mass behavior, below the $2\pi$ threshold, of the $N^* \to N$ transition form factor is of particular interest here. Faessler et al.~\cite{Faessler:2000md}, noting the too large values obtained for the radiative decay of nucleon resonances, had pointed out that this behavior could be cured by adding radial excitations of the $\rho$ meson. In effect, such addition changes the power of the pion form factor from monopole to dipole (or higher). Also the authors of Ref.~\cite{Larionov:2020fnu} had mentioned that the simple VMD2 version of Eq.~(\ref{VMD2}) is doubtful at small $m$.

In Ref.~\cite{HADES:2022vus} the discrepancy of the measured dilepton invariant mass distribution with the results of the VMD2 model (blue curve in Fig.~\ref{fig:invmassdileptons}) has been interpreted as a failure of the VMD2 version of vector meson dominance. It was indeed shown that agreement with the data could be obtained by using VMD1 of vector meson dominance and tuning the mixing parameters, that relate the direct photon-coupling amplitude to that for the $\rho$ coupling,  to the data \cite{Zetenyi:2020kon}. Alternatively, also a transition form factor model developed by Ramalho and Pena \cite{Ramalho:2016zgc} worked quite well. The latter, based on a quark-core and pion-cloud model, was actually predicted before the data became available.

Both of these models used an extrapolation of the $\rho$ spectral function into the region below the $2\pi$ threshold. Below the $2\pi$ threshold, however, the $\rho$ meson production cross section has not been measured. The cross section there is a convolution of the $\rho$ spectral function $A(m)$ and the dilepton decay width
\begin{equation}
	\frac{d\sigma}{dm} \propto A(m) \,\Gamma_{\rho \to e^+e^-}   ~.
\end{equation}
Any discrepancy for $m < 2m_\pi$ must, therefore, be attributed not necessarily to the dilepton decay width, but can also be due to a change in the spectral function $A(m)$ which in this mass region has not been measured.

 On the other hand, using the Breit-Wigner distribution of the $\rho$ meson also in the subthreshold region (as used in the GiBUU calculations, but also in the experimental analysis) then requires to change the Dalitz decay of the N(1520) resonance. As mentioned already earlier, at the end of Sect.~1, the $N(1520) \to N \rho$ decay vertex contains only a phase-space factor, the Blatt-Weisskopf formfactor does not appear in this case because the transition is an $s$-wave process. In order to account for a finite size of the hadronic transition one could thus introduce an additional formfactor which limits the population of high-momentum, very-low-mass $\rho$'s.
We note that similar formfactors appear also in an application of VMD1, both at the $N^*N\rho$ vertex and the $N^*N\gamma$ vertex.

The additional form factor that multiplies the decay width of the N(1520) resonance affects the population of the $\rho$ at small masses. There it can as well be seen as a multiplicative factor to the dilepton decay width.
In effect then this factor just makes the constant $C_\rho$ in Eq.~(\ref{VMD2}) mass dependent.

\section{Conclusions}
By comparing with a model (GiBUU) which contains both Fermi-motion and nuclear binding we have shown that the isolation of dilepton spectra on H out of data obtained for CH$_2$ and C is reasonably reliable. In addition the identification of the $\rho$ component in the dilepton spectrum, achieved by missing mass cuts, works well enough. In the absence of a realistic acceptance filter, however, quantitative differences between the normalization of the C and the H data still remain.

The obtained dilepton spectra deviate from the naive prediction of the vector meson dominance  model VMD2. While the authors of \cite{HADES:2022vus} have taken this as evidence against the validity of VMD2 in the present paper we have pointed out that these data can also be explained by either a formfactor at the $N^* N \rho$ vertex or, equivalently, by a mass dependence of the constant $C$ in the dilepton spectrum with still using VMD2. This explanation, as well as others based on VMD1, assumes that the Breit-Wigner distribution is still valid at very (unmeasured) low masses, far away from the peak mass. Alternatively the data could also be explained -- within VMD2 -- by a change in the spectral function of the $\rho$ meson at these low masses.

It is thus clear that a distinction between VMD1 and VMD2 can only be made if the $\rho$ meson spectral function is experimentally known. In the HADES experiment this is true only in the region above the $2\pi$ decay threshold. It is there, however, where VMD1 and VMD2 hardly differ even in the tune shown in \cite{HADES:2022vus}.

Finally, we note that while the GiBUU calculation indeed ascribes the dilepton yield mostly to a decay of a $\rho$ meson, originating in the N(1520) decay to the nucleon, it is not at all clear if this is also the case in the experimental data. Indeed, the dileptons observed in the $\pi^- + p \to n + e^+ e^-$ reaction can originate in a number of different elementary processes involving different hadronic transition form factors (see Fig.~40 in Ref.~\cite{Perez-Y-Jorba:1977vwf}).

\begin{acknowledgments}
We are grateful for a close collaboration with Alexei Larionov in the early stage of this investigation.
We also gratefully acknowledge helpful discussions with Stefan Leupold, Volker Metag, Beatrice Ramstein and Piotr Salabura.
This work was supported by BMBF grant no.~05P21RGFCA.
\end{acknowledgments}

\bibliography{H-dilept.bib}

\begin{thebibliography}{16}%
\makeatletter
\providecommand \@ifxundefined [1]{%
 \@ifx{#1\undefined}
}%
\providecommand \@ifnum [1]{%
 \ifnum #1\expandafter \@firstoftwo
 \else \expandafter \@secondoftwo
 \fi
}%
\providecommand \@ifx [1]{%
 \ifx #1\expandafter \@firstoftwo
 \else \expandafter \@secondoftwo
 \fi
}%
\providecommand \natexlab [1]{#1}%
\providecommand \enquote  [1]{``#1''}%
\providecommand \bibnamefont  [1]{#1}%
\providecommand \bibfnamefont [1]{#1}%
\providecommand \citenamefont [1]{#1}%
\providecommand \href@noop [0]{\@secondoftwo}%
\providecommand \href [0]{\begingroup \@sanitize@url \@href}%
\providecommand \@href[1]{\@@startlink{#1}\@@href}%
\providecommand \@@href[1]{\endgroup#1\@@endlink}%
\providecommand \@sanitize@url [0]{\catcode `\\12\catcode `\$12\catcode
  `\&12\catcode `\#12\catcode `\^12\catcode `\_12\catcode `\%12\relax}%
\providecommand \@@startlink[1]{}%
\providecommand \@@endlink[0]{}%
\providecommand \url  [0]{\begingroup\@sanitize@url \@url }%
\providecommand \@url [1]{\endgroup\@href {#1}{\urlprefix }}%
\providecommand \urlprefix  [0]{URL }%
\providecommand \Eprint [0]{\href }%
\providecommand \doibase [0]{http://dx.doi.org/}%
\providecommand \selectlanguage [0]{\@gobble}%
\providecommand \bibinfo  [0]{\@secondoftwo}%
\providecommand \bibfield  [0]{\@secondoftwo}%
\providecommand \translation [1]{[#1]}%
\providecommand \BibitemOpen [0]{}%
\providecommand \bibitemStop [0]{}%
\providecommand \bibitemNoStop [0]{.\EOS\space}%
\providecommand \EOS [0]{\spacefactor3000\relax}%
\providecommand \BibitemShut  [1]{\csname bibitem#1\endcsname}%
\let\auto@bib@innerbib\@empty
\bibitem [{\citenamefont {Nambu}\ and\ \citenamefont
  {Sakurai}(1962)}]{Nambu:1962zz}%
  \BibitemOpen
  \bibfield  {author} {\bibinfo {author} {\bibfnamefont {Y.}~\bibnamefont
  {Nambu}}\ and\ \bibinfo {author} {\bibfnamefont {J.~J.}\ \bibnamefont
  {Sakurai}},\ }\href {\doibase 10.1103/PhysRevLett.8.79} {\bibfield  {journal}
  {\bibinfo  {journal} {Phys. Rev. Lett.}\ }\textbf {\bibinfo {volume} {8}},\
  \bibinfo {pages} {79} (\bibinfo {year} {1962})}\BibitemShut {NoStop}%
\bibitem [{\citenamefont {O'Connell}\ \emph {et~al.}(1997)\citenamefont
  {O'Connell}, \citenamefont {Pearce}, \citenamefont {Thomas},\ and\
  \citenamefont {Williams}}]{OConnell:1995nse}%
  \BibitemOpen
  \bibfield  {author} {\bibinfo {author} {\bibfnamefont {H.~B.}\ \bibnamefont
  {O'Connell}}, \bibinfo {author} {\bibfnamefont {B.~C.}\ \bibnamefont
  {Pearce}}, \bibinfo {author} {\bibfnamefont {A.~W.}\ \bibnamefont {Thomas}},
  \ and\ \bibinfo {author} {\bibfnamefont {A.~G.}\ \bibnamefont {Williams}},\
  }\href {\doibase 10.1016/S0146-6410(97)00044-6} {\bibfield  {journal}
  {\bibinfo  {journal} {Prog. Part. Nucl. Phys.}\ }\textbf {\bibinfo {volume}
  {39}},\ \bibinfo {pages} {201} (\bibinfo {year} {1997})},\ \Eprint
  {http://arxiv.org/abs/hep-ph/9501251} {arXiv:hep-ph/9501251} \BibitemShut
  {NoStop}%
\bibitem [{\citenamefont {Abou~Yassine}\ \emph {et~al.}(2022)\citenamefont
  {Abou~Yassine} \emph {et~al.}}]{HADES:2022vus}%
  \BibitemOpen
  \bibfield  {author} {\bibinfo {author} {\bibfnamefont {R.}~\bibnamefont
  {Abou~Yassine}} \emph {et~al.} (\bibinfo {collaboration} {HADES}),\
  }\href@noop {} {\  (\bibinfo {year} {2022})},\ \Eprint
  {http://arxiv.org/abs/2205.15914} {arXiv:2205.15914 [nucl-ex]} \BibitemShut
  {NoStop}%
\bibitem [{\citenamefont {Adamczewski-Musch}\ \emph {et~al.}(2020)\citenamefont
  {Adamczewski-Musch} \emph {et~al.}}]{HADES:2020kce}%
  \BibitemOpen
  \bibfield  {author} {\bibinfo {author} {\bibfnamefont {J.}~\bibnamefont
  {Adamczewski-Musch}} \emph {et~al.} (\bibinfo {collaboration} {HADES}),\
  }\href {\doibase 10.1103/PhysRevC.102.024001} {\bibfield  {journal} {\bibinfo
   {journal} {Phys. Rev. C}\ }\textbf {\bibinfo {volume} {102}},\ \bibinfo
  {pages} {024001} (\bibinfo {year} {2020})},\ \Eprint
  {http://arxiv.org/abs/2004.08265} {arXiv:2004.08265 [nucl-ex]} \BibitemShut
  {NoStop}%
\bibitem [{{Bonn-Gatchina Partial Wave Analysis}()}]{BnGa}%
  \BibitemOpen
  {Bonn-Gatchina Partial Wave Analysis},\ \href@noop {} {}\bibinfo {note}
  {\url{http://pwa.hiskp.uni-bonn.de/}}\BibitemShut {NoStop}%
\bibitem [{\citenamefont {Anisovich}\ \emph {et~al.}(2005)\citenamefont
  {Anisovich}, \citenamefont {Klempt}, \citenamefont {Sarantsev},\ and\
  \citenamefont {Thoma}}]{Anisovich:2004zz}%
  \BibitemOpen
  \bibfield  {author} {\bibinfo {author} {\bibfnamefont {A.}~\bibnamefont
  {Anisovich}}, \bibinfo {author} {\bibfnamefont {E.}~\bibnamefont {Klempt}},
  \bibinfo {author} {\bibfnamefont {A.}~\bibnamefont {Sarantsev}}, \ and\
  \bibinfo {author} {\bibfnamefont {U.}~\bibnamefont {Thoma}},\ }\href
  {\doibase 10.1140/epja/i2004-10125-6} {\bibfield  {journal} {\bibinfo
  {journal} {Eur. Phys. J. A}\ }\textbf {\bibinfo {volume} {24}},\ \bibinfo
  {pages} {111} (\bibinfo {year} {2005})},\ \Eprint
  {http://arxiv.org/abs/hep-ph/0407211} {arXiv:hep-ph/0407211} \BibitemShut
  {NoStop}%
\bibitem [{\citenamefont {Larionov}\ \emph {et~al.}(2021)\citenamefont
  {Larionov}, \citenamefont {Mosel},\ and\ \citenamefont {von
  Smekal}}]{Larionov:2020fnu}%
  \BibitemOpen
  \bibfield  {author} {\bibinfo {author} {\bibfnamefont {A.~B.}\ \bibnamefont
  {Larionov}}, \bibinfo {author} {\bibfnamefont {U.}~\bibnamefont {Mosel}}, \
  and\ \bibinfo {author} {\bibfnamefont {L.}~\bibnamefont {von Smekal}},\
  }\href {\doibase 10.1103/PhysRevC.102.064913} {\bibfield  {journal} {\bibinfo
   {journal} {Phys. Rev. C}\ }\textbf {\bibinfo {volume} {102}},\ \bibinfo
  {pages} {064913} (\bibinfo {year} {2021})},\ \Eprint
  {http://arxiv.org/abs/2009.11702} {arXiv:2009.11702 [nucl-th]} \BibitemShut
  {NoStop}%
\bibitem [{\citenamefont {Buss}\ \emph {et~al.}(2012)\citenamefont {Buss},
  \citenamefont {Gaitanos}, \citenamefont {Gallmeister}, \citenamefont {van
  Hees}, \citenamefont {Kaskulov} \emph {et~al.}}]{Buss:2011mx}%
  \BibitemOpen
  \bibfield  {author} {\bibinfo {author} {\bibfnamefont {O.}~\bibnamefont
  {Buss}}, \bibinfo {author} {\bibfnamefont {T.}~\bibnamefont {Gaitanos}},
  \bibinfo {author} {\bibfnamefont {K.}~\bibnamefont {Gallmeister}}, \bibinfo
  {author} {\bibfnamefont {H.}~\bibnamefont {van Hees}}, \bibinfo {author}
  {\bibfnamefont {M.}~\bibnamefont {Kaskulov}},  \emph {et~al.},\ }\href
  {\doibase 10.1016/j.physrep.2011.12.001} {\bibfield  {journal} {\bibinfo
  {journal} {Phys.Rept.}\ }\textbf {\bibinfo {volume} {512}},\ \bibinfo {pages}
  {1} (\bibinfo {year} {2012})},\ \Eprint {http://arxiv.org/abs/1106.1344}
  {arXiv:1106.1344 [hep-ph]} \BibitemShut {NoStop}%
\bibitem [{{GiBUU Website}()}]{gibuu}%
  \BibitemOpen
  {GiBUU Website},\ \href@noop {} {}\bibinfo {note}
  {\url{http://gibuu.hepforge.org}}\BibitemShut {NoStop}%
\bibitem [{\citenamefont {Manley}\ and\ \citenamefont
  {Saleski}(1992)}]{Manley:1992yb}%
  \BibitemOpen
  \bibfield  {author} {\bibinfo {author} {\bibfnamefont {D.~M.}\ \bibnamefont
  {Manley}}\ and\ \bibinfo {author} {\bibfnamefont {E.~M.}\ \bibnamefont
  {Saleski}},\ }\href {\doibase 10.1103/PhysRevD.45.4002} {\bibfield  {journal}
  {\bibinfo  {journal} {Phys. Rev.}\ }\textbf {\bibinfo {volume} {D45}},\
  \bibinfo {pages} {4002} (\bibinfo {year} {1992})}\BibitemShut {NoStop}%
\bibitem [{\citenamefont {Post}\ \emph {et~al.}(2004)\citenamefont {Post},
  \citenamefont {Leupold},\ and\ \citenamefont {Mosel}}]{Post:2003hu}%
  \BibitemOpen
  \bibfield  {author} {\bibinfo {author} {\bibfnamefont {M.}~\bibnamefont
  {Post}}, \bibinfo {author} {\bibfnamefont {S.}~\bibnamefont {Leupold}}, \
  and\ \bibinfo {author} {\bibfnamefont {U.}~\bibnamefont {Mosel}},\ }\href
  {\doibase 10.1016/j.nuclphysa.2004.05.016} {\bibfield  {journal} {\bibinfo
  {journal} {Nucl. Phys. A}\ }\textbf {\bibinfo {volume} {741}},\ \bibinfo
  {pages} {81} (\bibinfo {year} {2004})},\ \Eprint
  {http://arxiv.org/abs/nucl-th/0309085} {arXiv:nucl-th/0309085} \BibitemShut
  {NoStop}%
\bibitem [{\citenamefont {Baldini}\ \emph {et~al.}(1988)\citenamefont
  {Baldini}, \citenamefont {Flaminio}, \citenamefont {Moorhead},\ and\
  \citenamefont {Morrison}}]{Schopper:1988vrx}%
  \BibitemOpen
  \bibfield  {author} {\bibinfo {author} {\bibfnamefont {A.}~\bibnamefont
  {Baldini}}, \bibinfo {author} {\bibfnamefont {V.}~\bibnamefont {Flaminio}},
  \bibinfo {author} {\bibfnamefont {W.~G.}\ \bibnamefont {Moorhead}}, \ and\
  \bibinfo {author} {\bibfnamefont {D.~R.~O.}\ \bibnamefont {Morrison}},\
  }\href {\doibase 10.1007/b33548} {\emph {\bibinfo {title} {{Total
  Cross-Sections for Reactions of High Energy Particles (Including Elastic,
  Topological, Inclusive and Exclusive Reactions)}}}},\ edited by\ \bibinfo
  {editor} {\bibfnamefont {H.}~\bibnamefont {Schopper}},\ \bibinfo {series}
  {Landolt-Boernstein - Group I Elementary Particles, Nuclei and Atoms}, Vol.\
  \bibinfo {volume} {12a}\ (\bibinfo  {publisher} {Springer},\ \bibinfo {year}
  {1988})\BibitemShut {NoStop}%
\bibitem [{\citenamefont {Faessler}\ \emph {et~al.}(2003)\citenamefont
  {Faessler}, \citenamefont {Fuchs}, \citenamefont {Krivoruchenko},\ and\
  \citenamefont {Martemyanov}}]{Faessler:2000md}%
  \BibitemOpen
  \bibfield  {author} {\bibinfo {author} {\bibfnamefont {A.}~\bibnamefont
  {Faessler}}, \bibinfo {author} {\bibfnamefont {C.}~\bibnamefont {Fuchs}},
  \bibinfo {author} {\bibfnamefont {M.~I.}\ \bibnamefont {Krivoruchenko}}, \
  and\ \bibinfo {author} {\bibfnamefont {B.~V.}\ \bibnamefont {Martemyanov}},\
  }\href {\doibase 10.1088/0954-3899/29/4/302} {\bibfield  {journal} {\bibinfo
  {journal} {J. Phys. G}\ }\textbf {\bibinfo {volume} {29}},\ \bibinfo {pages}
  {603} (\bibinfo {year} {2003})},\ \Eprint
  {http://arxiv.org/abs/nucl-th/0010056} {arXiv:nucl-th/0010056} \BibitemShut
  {NoStop}%
\bibitem [{\citenamefont {Z\'et\'enyi}\ \emph {et~al.}(2021)\citenamefont
  {Z\'et\'enyi}, \citenamefont {Nitt}, \citenamefont {Buballa},\ and\
  \citenamefont {Galatyuk}}]{Zetenyi:2020kon}%
  \BibitemOpen
  \bibfield  {author} {\bibinfo {author} {\bibfnamefont {M.}~\bibnamefont
  {Z\'et\'enyi}}, \bibinfo {author} {\bibfnamefont {D.}~\bibnamefont {Nitt}},
  \bibinfo {author} {\bibfnamefont {M.}~\bibnamefont {Buballa}}, \ and\
  \bibinfo {author} {\bibfnamefont {T.}~\bibnamefont {Galatyuk}},\ }\href
  {\doibase 10.1103/PhysRevC.104.015201} {\bibfield  {journal} {\bibinfo
  {journal} {Phys. Rev. C}\ }\textbf {\bibinfo {volume} {104}},\ \bibinfo
  {pages} {015201} (\bibinfo {year} {2021})},\ \Eprint
  {http://arxiv.org/abs/2012.07546} {arXiv:2012.07546 [nucl-th]} \BibitemShut
  {NoStop}%
\bibitem [{\citenamefont {Ramalho}\ and\ \citenamefont
  {Pe\~na}(2017)}]{Ramalho:2016zgc}%
  \BibitemOpen
  \bibfield  {author} {\bibinfo {author} {\bibfnamefont {G.}~\bibnamefont
  {Ramalho}}\ and\ \bibinfo {author} {\bibfnamefont {M.~T.}\ \bibnamefont
  {Pe\~na}},\ }\href {\doibase 10.1103/PhysRevD.95.014003} {\bibfield
  {journal} {\bibinfo  {journal} {Phys. Rev. D}\ }\textbf {\bibinfo {volume}
  {95}},\ \bibinfo {pages} {014003} (\bibinfo {year} {2017})},\ \Eprint
  {http://arxiv.org/abs/1610.08788} {arXiv:1610.08788 [nucl-th]} \BibitemShut
  {NoStop}%
\bibitem [{\citenamefont {Perez-Y-Jorba}\ and\ \citenamefont
  {Renard}(1977)}]{Perez-Y-Jorba:1977vwf}%
  \BibitemOpen
  \bibfield  {author} {\bibinfo {author} {\bibfnamefont {J.~P.}\ \bibnamefont
  {Perez-Y-Jorba}}\ and\ \bibinfo {author} {\bibfnamefont {F.~M.}\ \bibnamefont
  {Renard}},\ }\href {\doibase 10.1016/0370-1573(77)90044-8} {\bibfield
  {journal} {\bibinfo  {journal} {Phys. Rept.}\ }\textbf {\bibinfo {volume}
  {31}},\ \bibinfo {pages} {1} (\bibinfo {year} {1977})}\BibitemShut {NoStop}%
\end{thebibliography}%

\end{document}